\documentclass{ar-1col}

\usepackage[comma]{natbib}
\usepackage{url}
\usepackage{verbatim, amsmath, amsfonts, amssymb,amssymb}
\usepackage{aas_macros}
\usepackage{lipsum}
\usepackage{siunitx}
\usepackage{hyperref}
\usepackage{natbib}

\jname{Xxxx. Xxx. Xxx. Xxx.}
\jvol{AA}
\jyear{YYYY}
\doi{10.1146/((please add article doi))}

\begin{document}

\markboth{Feigelson et al.}{Challenges in Astrostatistics}

\title{21\textsuperscript{st} Century Statistical and Computational Challenges in Astrophysics}

\author{Eric D. Feigelson$^{1,2,3}$,  Rafael S. de Souza$^{4}$, Emille E. O. Ishida$^{5}$, and Gutti Jogesh Babu$^{2,1,3}$
\affil{$^1$ Department of Astronomy \& Astrophysics, Penn State University, University Park PA, USA, 16802, e5f@psu.edu}
\affil{$^2$ Department of Statistics, Penn State University, University Park PA, USA, 16802}
\affil{$^3$ Center for Astrostatistics, Penn State University, University Park PA, USA, 16802}
\affil{$^4$ Key Laboratory for Research in Galaxies and Cosmology, Shanghai Astronomical Observatory, Chinese Academy of Sciences, 80 Nandan Road, Shanghai 200030, China} 
\affil{$^5$ Universit\'e Clermont Auvergne, CNRS/IN2P3, LPC, F-63000 Clermont-Ferrand, France}
}

\begin{abstract}
Modern astronomy has been rapidly increasing our ability to see deeper into the universe, acquiring enormous samples of cosmic populations.  Gaining  astrophysical insights from these datasets requires a wide range of sophisticated statistical and machine learning methods.  Long-standing problems in cosmology include characterization of galaxy clustering and estimation of galaxy distances from photometric colors.  Bayesian inference, central to linking astronomical data to nonlinear astrophysical models, addresses problems in solar physics, properties of star clusters, and exoplanet systems. Likelihood-free methods are growing in importance.  Detection of faint signals in complicated noise is needed to find periodic behaviors in stars and detect explosive gravitational wave events. Open issues concern treatment of heteroscedastic measurement errors and understanding probability distributions characterizing astrophysical systems.  The field of astrostatistics needs increased collaboration with statisticians in the design and analysis stages of research projects, and to jointly develop new statistical methodologies. Together, they will draw more astrophysical insights into astronomical populations and the cosmos itself.
\end{abstract}

\begin{keywords}
astronomy, astrophysics, astrostatistics, cosmology, galaxies, stars, exoplanets, gravitational waves, Bayesian inference, likelihood-free modeling, signal detection, periodic time series, machine learning, measurement errors
\end{keywords}
\maketitle

\tableofcontents

\section{ASTRONOMY, ASTROPHYSICS, AND ASTROSTATISTICS} \label{intro.sec}

Astronomy, the oldest science, has had profound relationships to statistical concepts since antiquity.  Hipparchus chose the midrange estimator to reconcile inconsistent measurements of the duration of a year; Ptolemy and al-Biruni instead chose the mean value in their celestial calculations \citep{Sheynin:73}.  Galileo outlined a theory of observational errors in a discussion of the Comet of 1572. Newton had little interest in probabilistic arguments but his followers developing celestial mechanics laid the foundations of modern statistics.  Laplace minimized both the sum of absolute and squared residuals between observations and model predictions, but the $L_2$ method was integrated to the Gaussian error distribution in least squares theory \citep{Stigler:86}.  Many leading astronomers contributed to least squares methodology during the 19\textsuperscript{th} century.

But the links between astronomy and statistics sundered during the 20\textsuperscript{th} century as the former turned to physics and the latter to human affairs.  Statistician involvement in astronomy was rare and unsuccessful. \citet{Pearson08} presented a study quantifying a strong correlation between the range of variability and brightness of variable stars, but it was subject to  confusion between observed brightness and intrinsic luminosity and the mixing of pulsating and eclipsing variables.  Decades later, \citet{Neyman1952,Neyman1958} developed a model for galaxy clustering involving power law clusters placed at seed locations obtained by a Poisson point process.  However, it proved to be much too simple to describe for the hierarchical anisotropic spatial distribution of galaxies that emerged later from redshift surveys \citep{deLapparent86}.  

The divorce of astronomers from established methodology was at times severe.  \citet{Schlesinger10} had to plead with his community that least squares solutions to spectroscopic binary star orbits had a stronger foundation than approximate subjective methods.  \citet{Hubble30} obtained fits of elliptical galaxy light distributions to a spherical self-gravitating model by trial-and-error, and \citet{Zwicky37} made the seminal discovery of Dark Matter in the Coma cluster of galaxies with a curve fitted by eye.  The first widely recognized application of maximum likelihood estimation  did not appear until \citet{LyndenBell1988} applied it to a dynamical model of galaxy redshifts, discovering a previously unrecognized nearby concentration of galaxies known as the Great Attractor. 

Connections between the fields began to reestablish themselves in the 1990s and have rapidly grown.  In the astronomy research literature, terms like ‘Bayesian’ and ‘machine learning’ have increased exponentially in the past decade, with ‘Deep Learning’ rising meteorically since 2017.  While the number of professional statisticians working on astronomical problems is still small, the interest in advanced methodology $-$ particularly machine learning $-$ in the astronomical community is very strong.   

Observational astronomy today constitute a considerable enterprise with billions of dollars supporting $\sim$20,000 scientists producing $\sim$15,000 refereed papers annually. The science is devoted to the characterization and understanding of phenomena outside of Earth: our Sun and Solar System, other stars and their planetary systems, the Milky Way Galaxy and other galaxies, diffuse material between the stars and galaxies, and the Universe as a whole.  Progress is propelled by rapid development of technologies improving our observations at wavelengths from the longest radio waves to the shortest gamma-rays. Not all discoveries involve electromagnetic waves. Telescopes with strange designs detect energetic particles like cosmic rays and neutrinos, and most recently gravitational waves in space-time itself, give unique insights into explosive phenomena across the Universe.   

Theoretical astrophysics seeks to interpret telescopic results using physical processes known from terrestrial studies.  An astonishing range of physics is involved in cosmic phenomena: gravitational physics and fundamental theory; atomic and nuclear physics; thermodynamics, hydrodynamics and magnetohydrodynamics; molecular and solid state physics. Many observations cannot be closely linked to physics and are interpreted using heuristic statistical models like linear regressions;  these are often power law relationships because variables are plotted with logarithmic transformation.

But in other cases, convincing physical explanations are available and data are used to find best-fit parameters of complicated nonlinear astrophysical models.  One of the most important and successful astrophysical models in recent decades is the $\Lambda$CDM cosmological model: the expanding Universe with attractive cold Dark Matter and repulsive Dark Energy.  It gives a fundamental understanding of the evolution of the Universe from the Big Bang 13.7 billion years ago to the present day and into the future.  Ordinary matter of stars, planets and people constitute only a small fraction of the `stuff' in the Universe that is dominated by enigmatic, invisible material and forces.  An indicator of the powerful links between astronomy and physics are the eleven Nobel Prizes in Physics awarded for astrophysics in the past 50 years.  

While still a smaller enterprise than astrophysics, astrostatistics is playing an increasing role in the analysis of astronomical observations and linking data to astrophysical theory.  Consider the field of time domain astronomy.  While the stars seen with the naked eye seem unchanging throughout our lives, in fact many objects have variable characteristics  from exoplanets orbiting nearby stars to accreting black holes in distant quasars. Enormous investment in telescopes for repeated measurements over time are made, such as the Vera C. Rubin Observatory now under construction in the high Atacama Desert of Chile. Its main task will be an astronomical survey, the Legacy Survey of Space and Time (LSST), essentially a decade-long `movie' of variable objects in the sky. 

In this review, we present a non-comprehensive selection of issues important to current understanding of cosmic phenomena where progress seems impossible without sophisticated statistical analysis.  In some cases, astrostatisticians have had considerable success with established methods. In other cases, new developments are underway or the problems need creative ideas.  Our approach complements the more integrative review of \citet{Schafer15}.  Although our treatment is very incomplete, we hope to communicate to statisticians the unusual intellectual culture of astronomy with its amazing instruments, rapid progress, fascinating science, and exciting methodological challenges.

\section{TWO LONG-STANDING PROBLEMS IN COSMOLOGY}

\subsection{Galaxy Clustering}   \label{galclus.sec}

\begin{figure}[b]
\includegraphics[width=0.80\textwidth]{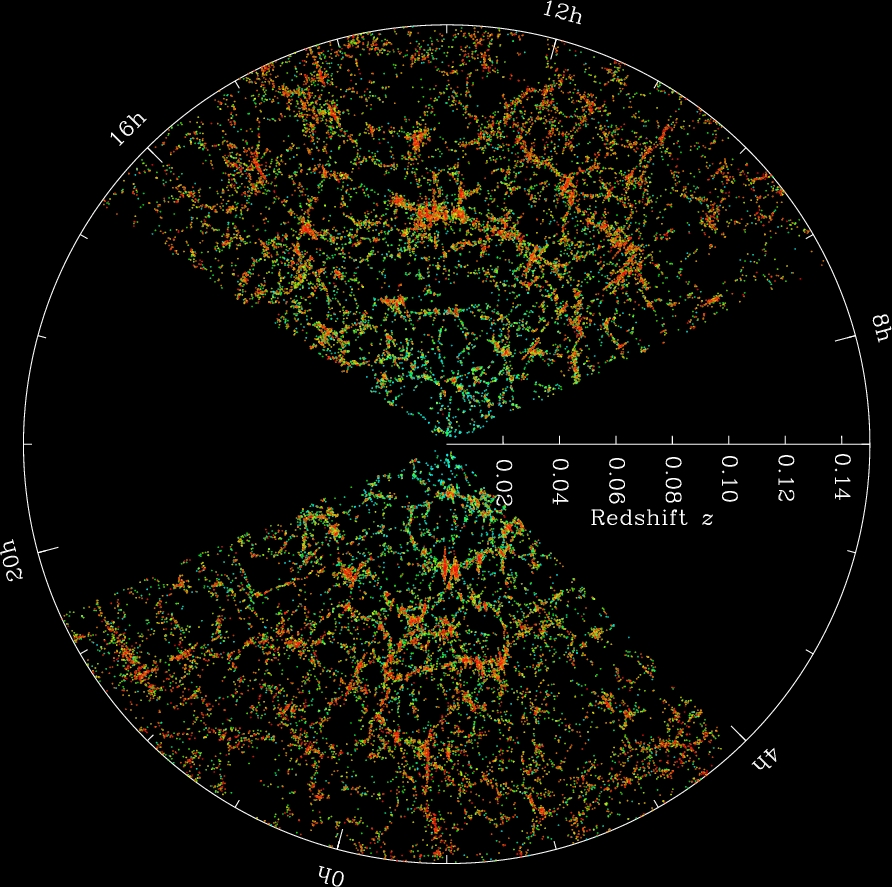}
\caption{Observed large scale structure from the Sloan Digital Sky Survey main galaxy redshift sample. The slice is 2.5$^\circ$ thick, and extends to redshift 0.14. Galaxies are color-coded by g-r color. Image from http://classic.sdss.org/legacy.}
\label{galaxy-lss.fig} 
\end{figure}

The distribution of galaxies in space proves to be surprising complex from the viewpoint of spatial point processes. Figure~\ref{galaxy-lss.fig} shows the distribution of galaxies in a slice of the sky out to redshift 0.14, equivalent to distances $\sim$600 megaparsecs (Mpc, a parsec is about 3.26 lightyears).  The dataset pictured here is part of the Sloan Digital Sky Survey (SDSS), one of the most successful astronomical projects of modern times that produced thousands of important studies.   

The galaxy distribution is very nonstationary and anisotropic on smaller scales ($< 200$ Mpc) but mostly stationary on large scales.  The pattern, commonly called the 'large-scale structure' of the Universe, roughly resembles a collection of contiguous soap bubbles where galaxies are distributed along curves `filaments' and sheets surrounding 'voids' \citep{Zel1982}.  Particularly prominent filaments are called `Great Walls'. Rich clusters appear at the intersections of filaments and are sometimes collected into `superclusters' that can include tens of thousands of galaxies.  The pattern is far more complex than statistically established stationary models like Neyman-Scott,  Mat\'ern or Cox processes \citep{Baddeley15}.  

\begin{figure}
\includegraphics[width=0.90\textwidth]{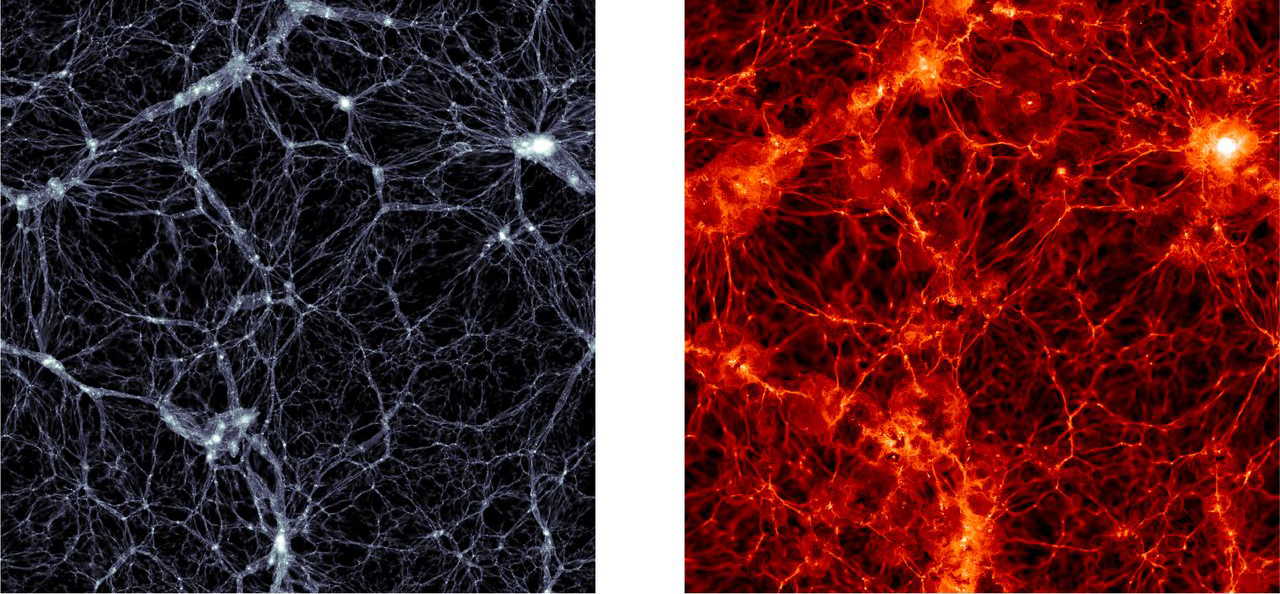}
\caption{Astrophysical simulation of large-scale structure.  A 2-dimensional slice today (13.7 billion years after the Big Bang) of a small portion of the 3-dimensional Illustris calculation of the growth of large-scale structure in the expanding Universe. The images show Dark Matter (left) and ordinary matter (right) density covering an area of $\sim 100 \times 100$~Mpc$^2$ \citep{Haider16}.}
\label{galaxy-lss-sim.fig} 
\end{figure}

Despite this complexity, the galaxy two-point (pair) correlation function is a simple power law (Pareto) function with a universal slope from 0.01 to 100 Mpc \citep{Peebles73}. This, however, does not prove to be a powerful discriminant between cosmological models. But the detection of a faint bump in the correlation function around 300 Mpc separation validated an important prediction of specific Big Bang theories.  Known as the baryonic acoustic oscillation signal, its discovery was one of the major achievements of the SDSS \citep{Eisenstein05}.  

Many statistical studies of large-scale structure rely on isotropic two- and three-point correlation functions as well as Fourier power spectra. But other studies seek to locate particular clusters, filaments or voids.  Several filament finding techniques have been investigated.  The pruned Minimal Spanning Tree is a popular procedure \citep{Barrow85} but the resulting filaments are often noisy with spurs and chaining. \citet{Stoica10} develop a model based on point processes marked by a filament identifier with probabilities favoring networks of multiply-connected aligned segments.  Global Bayesian solutions are found using a simulated annealing algorithm based on Metropolis-Hastings dynamics.  Other filament-finding algorithms are based on scalar measures of size and shape  \citep{Sahni98}, thresholded loci of density saddle points \citep{Novikov06}, segmented watershed transform \citep{Platen07}, measures based on the Hessian eigenvalues \citep{Bond10}, and density ridges traced by a subspace constrained mean shift algorithm \citep{Chen15,Moews2020}.  

Some of these algorithms are applied to the observed point spatial distribution and others are applied to its smoothed density estimator.  The latter situation also occurs when filamentary structures appear in continuous media traced in real-valued images rather than point processes.  Filaments are characteristic in hot flare plasma on the surface of the Sun and in cold molecular clouds within the Milky Way Galaxy.  \citet{Menshchikov10} defines molecular cloud filaments using the tint fill algorithm from image processing; they are astrophysically explained as outcomes of supersonic magnetohydrodynamic turbulence \citep{Beattie20}.  

In the case of galaxy clustering, powerful simulations of clustering patterns are available from astrophysical models involving local gravitational contraction within the expanding Universe following the Big Bang.  These are computationally intensive calculations with names like the {\it Millenium} \citep{Springel05} and {\it Illustris} \citep{Nelson15} simulations.  The models and observed galaxy clustering patterns are in considerable agreement, validating the dominance of Dark Matter in large-scale structure formation (Figure~\ref{galaxy-lss-sim.fig}).  Important issues regarding bias in visible galaxy formation and quasar feedback are still under investigation.  

But important methodological problems remain.  First, there has been little comparison of the performance of the various filament-finding methods or other measures of clustering.  Most are algorithmic with little foundation in mathematical or statistical theory.  Second, astronomers have few tools to compare complicated observed and simulated two- and three-dimensional point distributions with complexities, such as Figures~\ref{galaxy-lss.fig} and \ref{galaxy-lss-sim.fig}.  These tests must treat millions of points for quantitative comparison of model predictions and observed data.   

\subsection{The Photo-z Conundrum} \label{photoz.sec}

It is relatively easy in astronomy to measure locations on the celestial sphere but very difficult to measure distances.  Distances to stars in our region of the Milky Way Galaxy can be obtained from annual parallax measurements, a method known to Copernicus though not achieved until the 19$^{th}$ century. But distances to other galaxies must rely in indirect procedures.  For galaxies lying billions of parsecs away, estimating distances is paramount to understand their evolution across cosmic time. With the best telescopes, we can now see galaxies forming only $\sim 1$ Gyr after 
the Big Bang. But as the Universe is expanding, the spectrum is 
shifted towards longer (redder) observed wavelengths ($\lambda_{obs}$) compared to their rest-frame $\lambda_{rest}$ wavelengths.  The redshift is defined to be  $z = (\lambda_{obs}-\lambda_{rest}) / \lambda_{rest}$.  The astronomer obtains a spectrum of a galaxy with a spectrograph on a telescope to measure the wavelength in which 
specific spectral features are found. This is compared with the expected wavelength for that feature in rest frame.   For example, a galaxy with redshift $z=3$ has its hydrogen Balmer break shifted from the blue around 400 nm to the infrared around 1200 nm. 

Once the expansion rate of the Universe, known as Hubble's constant $H_0$, is measured, then the galaxy's distance in parsecs can be inferred from the redshift. After great effort over several decades,  $H_0$ is now known to be around 70 km/s/Mpc to within $\sim 2$\%.  Once the distance is obtained, then the galaxy's size, luminosity, mass, cluster environment, and stellar mass function are inferred from observed properties, as well as its chemical, merger and star formation history.  Obtaining accurate redshifts is thus essential for understanding the formation and properties of galaxies and to trace the evolution of large-scale structures (\S\ref{galclus.sec}). 

However, it is expensive in telescope time to obtain sufficiently high-resolution spectra in order to measure spectral lines and directly obtain redshifts.  It is much cheaper to measure brightnesses of many galaxies simultaneously in wide spectral bands; this is called photometry and produces a low-resolution 'spectral energy distribution' or SED. Here each band blurs together many spectral lines with continuum starlight.  A wide-field imaging camera may obtain 
photometry 100-fold more efficiently than a multi-object spectrograph for faint high-redshift galaxies \citep{Hildebrandt08}.  The redshift information is encoded in the photometry but not as directly as in a high-resolution spectrum.

Therefore, photometric redshift (photo-\emph{z}) estimation has become a  vital tool in the extragalactic astronomy and observational cosmology.  The challenge of photo-\emph{z} accuracy then depends on the statistical procedures used to calibrate 
photometric measurements to spectroscopic redshifts. 

A plethora of methods have been proposed and tested for this task \citep{benitez2000, beck2016B, Budavari2009, Elliott2015, Cavuoti16}. They range from several variants of least squares weighted by photometric measurement errors
to Bayesian inference \citep{Leistedt2016}, k-nearest neighbor procedures, kernel density estimation, Gaussian mixture models, generalized linear models, Self-Organizing Maps, Random Forest \citep{Carliles2010}, hyper-optimized gradient boosting regression, Gaussian processes regression, and hybrid schemes.  Galaxy morphology from images can supplement photometric measurements when neural networks are used.  More recently, some more advanced methods have been used, including the use of  different flavors of deep convolutional networks to derive  photometric redshift directly from multi-band images \citep{Isanto2018,Pasquet2019}, and to build a  morphology-aware photo-\emph{z} estimator \citep{Menou2019}. Comparisons of performance are presented by \citet{Dahlen13}, \citet{Rau15}, \citet{Salvato19}, and \citet{Schmidt20}. 

\begin{figure}
\includegraphics[width=0.99\textwidth]{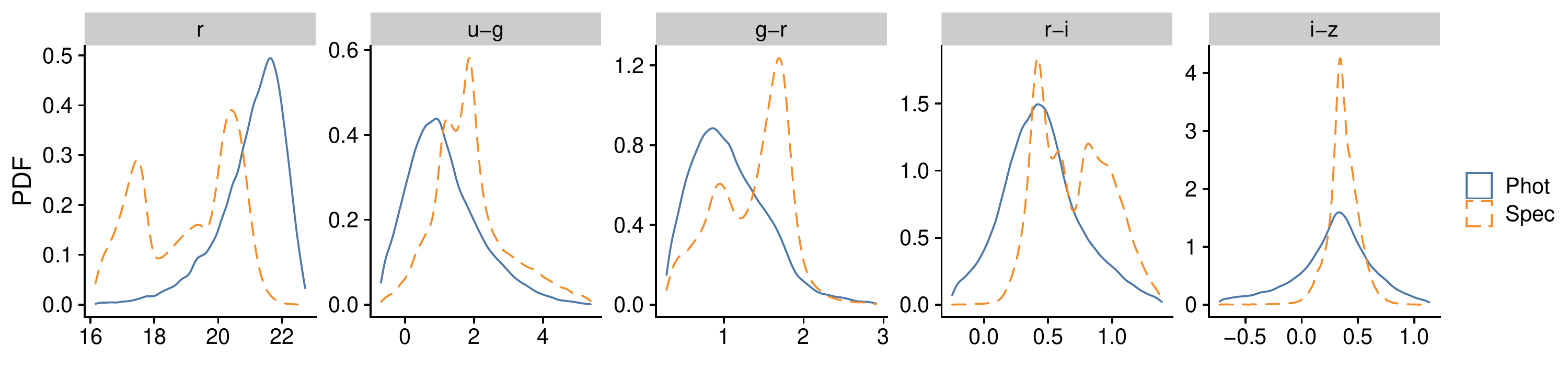}
\caption{Distributions for magnitude in r-band and four colors (u-g, g-r, r-i, i-z) for the spectroscopic (training)  sample compared with the photometric (target) sample in the data sets presented by \citet{Beck2017}.}
\label{photoz_dist.fig} 
\end{figure}

However most methods fail to achieve better than $\sim 2$\% accuracy. The relationship between photometry  
and spectroscopic redshifts are not only nonlinear, but degeneracies, heteroscedasticity, and `catastrophic outliers' abound in the datasets.  A further caveat encountered by machine learning based methods is a common mismatch between the parameter distribution of the training (spectroscopic) and target (photometric) datasets as portrayed in Figure~\ref{photoz_dist.fig}. This is connected with the nature of spectroscopic measurements which, within the same survey, demands higher quality data and consequently delivers lower redshift and higher metallicities galaxies then the photometric counterpart.

In cases where spectroscopic and photometric samples have similar coverage in magnitude/color space, it is feasible to adapt the spectroscopic sample using Domain Adaptation \citep{Beck2017}. However, in most real-data scenarios this assumption will not hold. An alternative  approach relies on active learning to improve the training set by making sensible decisions to query  extra galaxies from which one could measure the spectra \citep{Vilalta2017}.

\section{BAYESIAN MODELING OF THE SUN, STARS, AND PLANETARY SYSTEMS} \label{Bayesian.sec}

\subsection{Solar Magnetic Fields} \label{subsec:magfields}

\begin{figure}
    \includegraphics[width=0.99\textwidth]{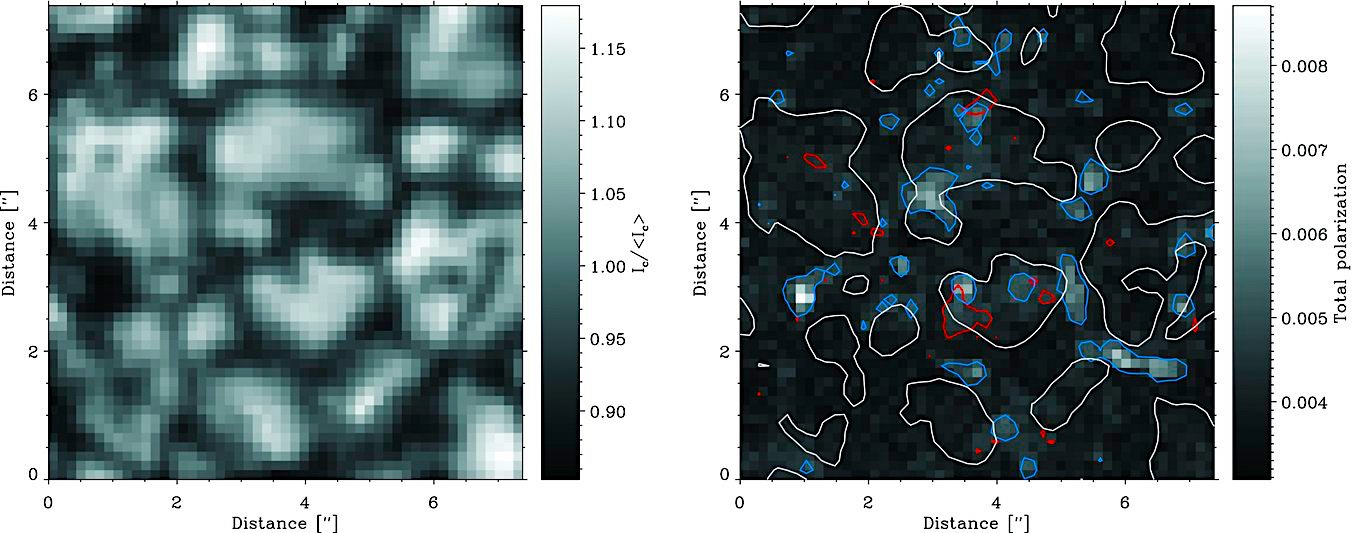}
    \includegraphics[width=0.99\textwidth]{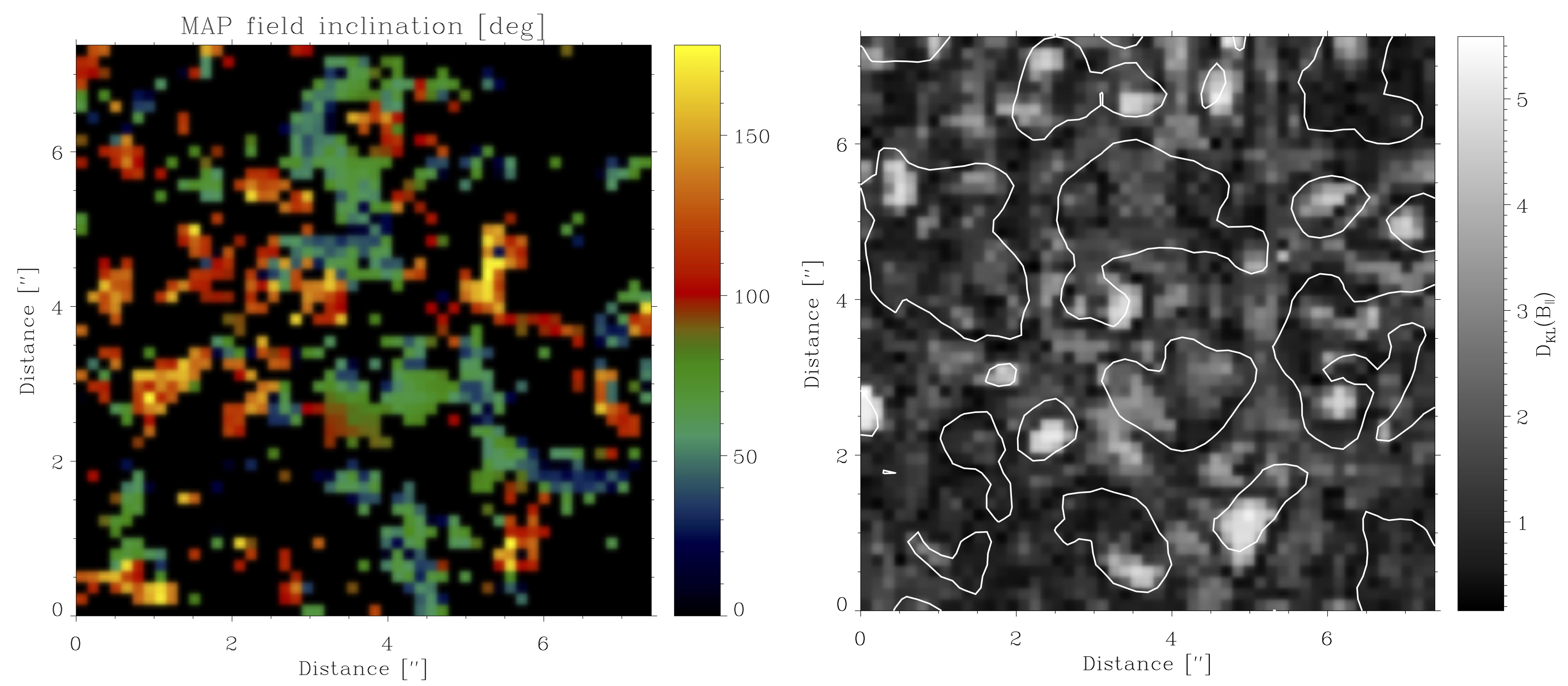} 
    \caption{Bayesian analysis of magnetic fields in the quiet Sun's photosphere \citep{Asensio2009}.  {\it Top left:} Map of the continuum brightness for a small region of the Sun.  {\it Top right:} Map of the polarization brightness where the white contours mark areas of bright continuum, red contours mark bright linear polarization, and blue contours mark bright circular polarization.  {\it Bottom left:} Maximum-a-posterior map of the magnetic field inclination inferred from the Bayesian model.  {\it Bottom right:} Map of the Kullback-Leibler divergence between prior and posterior for the magnetic field inclination. }
    \label{solar.fig}
\end{figure}

Studies of the Sun are among the active fields of astrophyical modeling, particularly addressing the many manifestations of its magnetic field generated by gas motions in its interior.  \citet{Asensio2006} presents a Bayesian model for the spatial variations of linear and circular polarization which can be linked to the underlying magnetic field strength through physical processes like radiative transfer and the Zeeman effect.  The method was applied to a polarization map of a small portion of the quiet Sun (Figure~\ref{solar.fig}) to investigate the role of magnetic fields in granulation arising from convection in its upper layers. The calculation is made for each pixel in the map, and maps are constructed of the Kullback-Liebler divergence between the prior and posterior distributions for several physical quantities (such as the magnetic field strength, filling factor, and geometry) after marginalization over other parameters. The resulting maps are similar to those obtained by traditional weighted least squares calculations (designated `$\chi^2$ minimization' in the astronomical community) but with improved values for magnetic field strength and treatment of degeneracies between the physical parameters.   

\subsection{Star Cluster Properties} \label{subsec:starcluster}

Over a century ago, it was discovered that most stars lie along a curved locus in the Hertzsprung-Russell (H-R) diagram, a plot of log-luminosity against surface temperature or color.  As nuclear physics developed, astrophysical models were  constructed showing that this 'main sequence' represents the long-lived phase powered by the fusion of hydrogen to helium in the stellar cores.  When hydrogen is depleted the cores, the stars move off the main sequence to higher luminosities and redder surface colors, the `red giant' regime.  The theory of stellar evolution has been well-developed for decades; the evolution of a star in the H-R diagram can be calculated based on nuclear, gravitational, atomic and fluid physics for a given mass and abundance of elements,  summarized in a parameter called `metallicity' (Z). Astronomers now use the evolutionary models to infer precise stellar properties that are otherwise difficult to ascertain, such as their ages ($\tau$) and masses (M).   

A long-standing astrostatistical collaboration has applied Bayesian inference to this classical problem in stellar astronomy \citep{vandyk2009}.  The data are easily obtained  from photometric measurements of the brightness $B_{ij}$ in several color bands $j$ for $i=1, ..., N$ stars in a coeval star cluster, together with measurement errors $\sigma(B_{ij})$. The likelihood is a product of Gaussians, 
\begin{equation}
    \mathcal{L}(M, \tau, Z |B, \sigma(B)) ~ \propto ~ \prod_{j=1}^k \prod_{i=1}^N \exp{\left[\frac{-[B_{ij}-G_j(M, \tau, Z)]^2}{2 \sigma^2({B_{ij}})}\right]},
\end{equation}
where $G_j$ are the predicted brightness from astrophysical stellar evolution models. Astronomers have an informative prior distribution for stellar masses;  stellar Initial Mass Function (IMF) is known to follow a log-normal distribution for stars in the range $0.1-8$ solar masses (\S\ref{statdistr.sec}). However, there is less knowledge about the star formation history of the Galaxy, so an uninformative uniform prior in log-age over a wide range is assumed. Gaussian priors are given to other stellar parameters such as metallicity.  Posterior values are calculated by Markov chain Monte Carlo using a one-at-a-time Gibbs sampler with a uniform Metropolis jumping rule.  This is a high-dimensional model with $3N+5$ parameters; a large look-up table of $G$ values is used to reduce stellar model calculations.  More complex likelihood models can be used to account for contamination by irrelevant field stars or for multiple age components in the star cluster.

The method is now widely used to estimate ages of star clusters in the Milky Way galaxy.  \citet{Bossini2019}  find ages ranging from 10 to 10,000 million years for hundreds of star clusters newly identified from the Gaia catalog of 1.3 billion Galactic stars.  The estimated masses agree very closely to values obtained by a more accurate method (modeling of asteroseismological  oscillations) available for a handful of stars.  New star clusters have been recently  discovered \citep{Cantat2019}, and there are many more to come. 

\subsection{Modeling Multiplanet Systems} \label{subsec:multiplanet}

One of the great excitements of modern astronomy is the clear evidence that most stars in the sky have their own planetary systems. The original discovery was based on radial velocity measurement of the host star by \citet{Mayor1995} who were consequently awarded the 2019 Nobel Prize in Physics.  The star is pulled back and forth as it is orbited by planets which are usually too close to the star to be resolved in a telescope image.  Spectra to obtain the star's radial velocity are  repeatedly obtained to detect the resulting periodic Doppler redshift and blueshift.  Extremely precise spectrographs are needed, as the velocity shifts may be less than 1 meter/sec, a tiny fraction of the velocity of light at 300 million meters/sec. 

The resulting dataset is a time series of radial velocity measurements.  Unfortunately, the cadence of observing times is sparse (as the spectra require considerable allocation of telescope time) and irregularly spaced (\S\ref{irregTS.sec}).  Furthermore, the observed Doppler shifts are subject to both instrumental noise and a `jitter' intrinsic to the star caused by magnetic activity similar to that seen in the Sun (\S\ref{subsec:magfields}).  

Fortunately, we have a reliable astrophysical model for the orbits of planets around a star, essentially the same that Newton's Laws gave for celestial mechanics in our own Solar System two centuries ago.  The parameters for the orbit of each planet include the period (in days), radial velocity amplitude (in meters/sec), eccentricity, four angles, a velocity calibration, and at least one parameter for the stellar jitter. With nine parameters for each planet, and a crucial additional parameter on the number of planets in the system, modeling a radial velocity time series involves optimizing a non-linear high dimensional problem. 

This problem is typically formulated in a Bayesian inferential framework \citep{Ford05}.   Choices of Bayesian priors and computational method have been carefully considered in light of astrophysical constraints \citep{Clyde2007, Sharma2017}.  With clever choice of variable transformations, MCMC method, and implementation on GPU or cloud supercomputers, the calculation can be successful.  An illustration of this approach, using a combination of radial velocity and transit measurements, is the discovery of a third planet orbiting Kepler-47 that gives new insights into planet formation processes and planet habitability around binary stars \citep{Orosz2019}.

\section{LIEKLIHOOD-FREE MODELING} \label{sec:ABC}

Two main forms of statistical models can be distinguished: those describe by probability distributions for which an explicit likelihood can be written; and implicit models, as with cosmological simulations (Figure~\ref{galaxy-lss-sim.fig}) from which one can simulate samples but without explicit likelihood formulations. The latter are often called generative models.  In astronomy, an important example of a simple linear parametric models is the  $M_{BH}$--$\sigma_{gal}$ relation that describes the empirical relation between the mass of the supermassive black hole present in the center of a galaxy and the velocity dispersion of the stars in its bulge \citep{Gebhardt00}. This is a heuristic model that does not (yet) derive from astrophysical insights.  

Approximate Bayesian Computation (ABC) enables parameter inference for complex physical systems in cases where the true likelihood function is unknown, unavailable, or computationally too expensive. It relies on the forward simulation of mock data and comparison between observed and synthetic results.  The underlying idea is to validate the theoretical model by confronting its simulated outcomes with those observed in the real world. A model which produces sufficiently realistic synthetic results is more likely to be close to the truth then another one which does not, as with measured and simulated large-scale structure properties (\S\ref{galclus.sec}). This concept is natural to astronomers; strategies very similar to -- and sometimes inspired by -- ABC were developed in the astronomical literature \citep{Lin2015b, Killedar2017}. 

Shortly after \citet{Schafer2012} highlighted the potential of ABC in astronomy, \citet{Cameron2012} used it to investigate the morphological evolution of distant galaxies. Since then the subject has attracted increasingly more attention including cosmological applications based on supernovae \citep{Weyant2013}, galaxy clusters number counts \citep{Ishida2015}, simulation images calibration \citep{Akeret2015} and weak lensing peak counts \citep{Lin2015b}.  ABC methods have also tackled subjects like galaxy star formation histories \citep{Hahn2017}, inter-galactic medium \citep{Davies2018}, exoplanetary systems \citep{Hsu2018}, accretion disks around supermassive black holes \citep{Witzel2018}, luminosity functions \citep{Riechers2019}, and the stellar IMF \citep{Cisewski2019}.
This movement is accompanied by the development of specialized software, designed by astronomers, which have played an important role in the dissemination of the technique such as  CosmoABC \citep{Ishida2015}, abcpmc \citep{Akeret2015}, and  astroABC \citep{Jennings2017}. 

As the astronomical community became more familiar with these statistical tools, it also became aware of its bottlenecks -- and started to propose solutions for crucial aspects the  paradigm. For example, the high number of simulations required by ABC can be prohibitive for many astronomical cases where calculations are computationally expensive. \citet{Kacprzak2018} proposes a quantile regression strategy which approximates the behavior of the distance function, thereby identifying regions of the parameter space with low probability to avoid generating samples dissimilar from the observed data. \citet{Alsing2018} propose a Density Estimation Likelihood-Free Inference algorithm where information from all available simulations are used to estimate a joint distribution of data and parameters, leading to a more efficient sampling then the traditional Population Monte Carlo approach. 

They also address the choice of summary statistics, proposing a 2-step algorithm that reduces dimensionality by combining heuristic summary statistics and Fisher information \citep{Alsing2018a}.  \citet{Charnock2018} suggest replacing the user defined summary by a neural network which is able to find nonlinear functions of the data that maximize Fisher information. \citet{Alsing2019} push the paradigm forward with neural density estimators to learn the likelihood function from a set of simulated data sets and coupling it with active learning \citep{Settles2012} to adaptively acquire simulations in the most relevant regions of parameter.

In less than 10 years, likelihood-free inference techniques like ABC and its  derivatives have flourished within the astronomical community. Given the high complexity of data expected from the next generation of large scale surveys, these efforts will prove valuable to the future of parametric inference in modern astronomical data.

\section{CHALLENGES IN SIGNAL DETECTION} \label{sec:signal_detection}

\subsection{Periodicity Detection in Irregular Time Series} \label{irregTS.sec}

The study of variable objects in the sky $-$ time domain astronomy $-$ is burgeoning with more than 2000 studies annually \citep{Griffin19}.  A common study involves the multi-epoch measurement of brightness of variable stars or quasars; astronomers call these time series `light curves'.  Yet, except for Fourier analysis and occasionally wavelet analysis pioneered by \citet{Starck1997}, astronomers rarely use statistical methods of time series analysis established for use in signal processing and econometrics (Box et al.\ 2015).  

The principal reason is that most multi-epoch programs have irregularly spaced cadences.  Some causes are unavoidable: for a single mountaintop telescope using visible light, a star or quasar is unobservable during daylight and is totally unobservable for half the year due to the annual solar motion. Other causes are sociological: telescope allocation committees must juggle projects by many scientists and can rarely give regularly spaced cadences. There are a few exceptions to these difficulties, such as NASA's Kepler satellite that reported the brightness of $\sim$200,000 stars every 30 minutes for four years, or the HAT South network of three small telescopes on different continents to reduce diurnal gaps. But irregular cadences are the norm in astronomical time series. 

As outlined in \S\ref{intro.sec}, cosmic objects exhibit an incredible variety of temporal characteristics in all parts of the electromagnetic spectrum, some of which are strictly periodic.  These might be a rapidly rotating neutron star with narrowly beamed radio emission (\S\ref{subsec:GW}) or an exoplanet transiting a single star with periodic variations in velocity and brightness (\S\ref{subsec:multiplanet}). These situations are sufficiently common and important that astronomers have developed a suite of methods for characterizing both periodic and aperiodic variations in irregular time series. However, their statistical foundations are often inadequate, and the quality of scientific inferences consequently are often unreliable.

Astronomers first developed nonparametric periodograms that provide signals in flexible time domain situations: irregular cadences; non-sinusoidal shapes (e.g., for brief transits); and heteroscedastic measurement errors (\S\ref{measerr.sec}).  The most widely used for variable star characterization is \citet{Stellingwerf78} `phase dispersion minimization' (PDM) periodogram.  Here the observations are folded modulo a trial period $P$, the data are grouped into a small number $k$ of evenly spaced bins, the weighted mean $x_i$ and standard deviation $\sigma_i$ for $i = 1, ..., k$ bins are obtained, and the PDM($P$) is the ratio of within-bin to between-bin dispersions weighted by measurement errors.  The periodogram consists of PDM values for a large number of trial periods.  If periodic behavior is present then, at the correct period $P_0$,  the high (low) brightnesses are collected in one or a few bins, and the PDM periodogram  has a dip at $P_0$. If there is no periodicity, then the high (low) brightnesses are distributed randomly among the bins and the PDM shows only noise values. Another formulation called the Analysis of Variance periodogram is mathematically related to the PDM  \citep{SchwarzenbergCzerny96}.

While PDM($P$) has a definable distribution with large sample Gaussian white noise, the conditions are often unfavorable. Too few data points may be present in some bins for accurate variance estimates; indeed, some bins can be empty for some trial periods. Scatter within a bin is often non-Gaussian with outliers from poorly quantified measurements. The choice of $k$ is arbitrary, the probabilities for False Alarms in the multiple trials has not been addressed, and the periodogram is subject to strong aliasing effects when true periods are present.  Perhaps the most insidious problem is that many stars show autocorrelated but aperiodic variations that interact with the irregular cadence to produce false peaks in the periodogram for short duration datasets.  None of these statistical issues have been examined by mathematical statisticians despite its use in $\sim 1500$ astronomical papers over four decades.  

Variants of epoch-folding periodograms have been developed. \citet{Dworetsky83} proposes an unbinned $L_1$ version of the PDM nicknamed the `minimum string length' periodogram that alleviates some problems, but in practice seems less sensitive than PDM. The detection of exoplanet transits, where the shape of the dip in brightness is box-shaped as the planet passes in front of the star, is commonly based on Box Least Squares regression procedure \citep{Kovacs02}.  \citet{Caceres19a} recently developed an efficient algorithm, nicknamed Transit Comb Filter, designed for a differenced transit light curve that calculates a matched filter to a periodic double-spike pattern rather than a periodic box pattern. 

The most commonly used tool for period searching in irregular cadence astronomical light curves is the Lomb-Scargle periodogram \citep[LSP,][with 4000 citations]{Scargle82} that assumes sinusoidal periodic behavior. It is a generalization of the Schuster periodogram in Fourier analysis for irregularly cadences. It often shows higher signal-to-noise and weaker aliases than PDM in tests with simulated data.   

But debates have waged on evaluating the statistical significance, or False Alarm Probability (FAP), of LSP peaks for realistic data.  The analytic exponential distribution of peak power applies only to idealized data: an infinite stream of regularly spaced Gaussian white noise with known variance and a single sinusoid superposed \citep{Percival93}. \citet{Koen1990} argues that the substitute of the sample variance for the population variance in this exponential formula can lead to badly underestimated errors in FAPs.  Several groups develop links between the LSP and Bayesian models, including an odds ratio for LSP peak significance \citep{Bretthorst03, Mortier15}.  LSP FAPs based on generalized extreme value (GEV) distributions have a strong performance in the analyses of \citet{Baluev08} and \citet{Suveges15}. \citet{Sulis17} combine bootstrap resampling with GEV distributions to estimate FAPs, and \citet{Delisle20} develop it further with computationally efficient approximations that work in the presence of correlated noise. GEV-based approaches have garnered some support in the community, but many practitioners are still using naive, unreliable FAPs.  In his review on understanding LSPa, \citet{VanderPlas18}  concludes:
\begin{quote} 
Unfortunately, there is no silver bullet for answering these broader, more relevant questions of uncertainty of Lomb-Scargle results. Perhaps the most fruitful path toward understanding of such effects for a particular set of observations—with particular noise characteristics and a particular observing window—is via simulated data injected into the detection pipeline. 
\end{quote}

\subsection{Gravitational Wave Detection} \label{subsec:GW}

\begin{figure}[h]
    \centering
    \includegraphics[scale=.13]{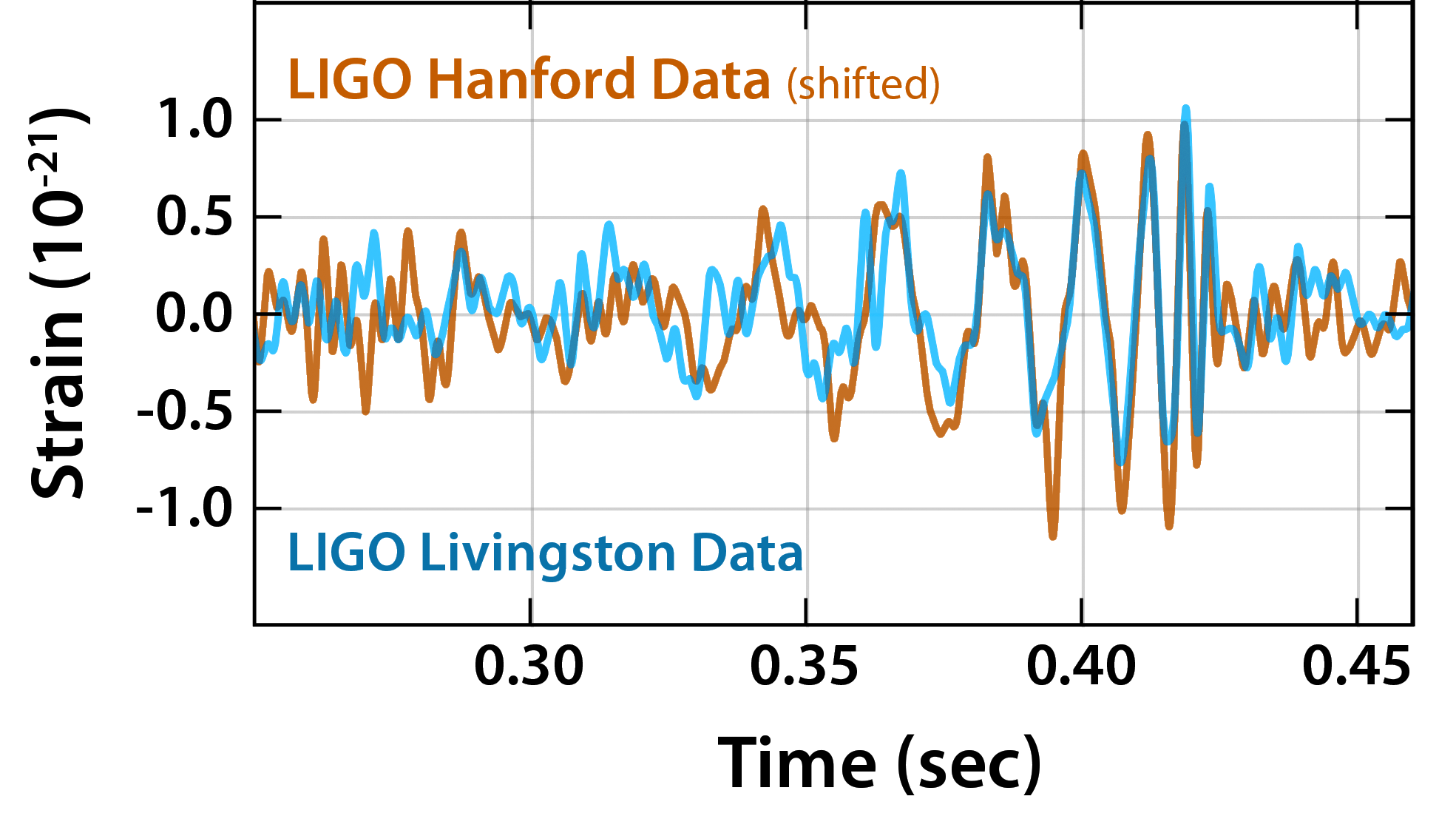}
    \caption{Discovery of the first gravitational wave event, GW15091, from the U.S. Laser Interferometric Gravitational Observatory. The plot shows a chirp signal from two widely separated interferometers after shifting by a short time delay. This event arises from the inspiraling and coalescence of two $\sim 30$ solar mass black holes in an unknown distant galaxy.  Credit: Caltech/MIT/LIGO Lab.
}    \label{fig:ligo150914}
\end{figure}

The dawn of a new era of astronomy came in 2015 with the detection of Gravitational Waves (GWs). A century earlier, Albert Einstein predicted that changes in gravitational fields would cause tiny ripples in space-time to propagate through the Universe.  The effect is so weak that only the most sophisticated instruments can detect the strongest gravitational events, such as the inspiralling and merger of two black holes or neutron stars (Figure~\ref{fig:ligo150914}). These remarkable events are rare, but a growing number have now been detected in other galaxies with the Laser Interferometer Gravitational-Wave Observatory (LIGO).  Other observatories are seeking GWs, such as ground-based radio pulsar timing arrays and planned space-based multi-satellite interferometers. This new field of astronomy, which garnered the 2017 Nobel Prize in Physics, will open new opportunities for confronting both astrophysical theories (binary star evolution, black hole formation, merging galaxies, quasar evolution) and fundamental physics (high density matter, particle physics, theory of gravitation) in ways that are otherwise inaccessible.  

The LIGO observatories (now in the U.S. and Italy, soon to expand in Japan and India), have laser beams of infrared light traveling along two perpendicular kilometer-scale vacuum chambers where mirrors and beam splitters allow tiny motions in the length of the device to be measured using interference patterns. LIGO is designed to detect GW events on timescales of $10^{-4}-10^{-1}$ seconds corresponding to violent changes in stellar mass bodies such as white dwarfs, neutron stars and black holes arising from supernova explosions. In some cases, the GW event is accompanied by a burst of gamma-rays or other electromagnetic radiation; this was first seen in GW170817 \citep{Abbott17}. Time delays between detections at the different observatories allow triangulation of the GW location in the celestial sphere, so other observatories can quickly search for electromagnetic counterparts. The scramble to detect signals from different observatories associated with GW and similar explosive events is nicknamed `multi-messenger astronomy'.

The statistical challenge with LIGO is to detect sudden short-lived chirp-like events in a continuous time series (Figure~\ref{fig:ligo150914}) where noise is dominated by instrumental effects that can be continuous (perhaps caused by  vibrations in the mirror structures) or transient (perhaps caused by minor Earth tremors). We seek brief, weak signals in non-stationary, non-Gaussian noise.  An elaborate procedure for removing or flagging these non-astrophysical variations has been developed \citep{McIver12, Cabero19}.  The resulting residual time series is nearly Gaussian white noise, and GW chirps are most often sought by matched filtering with astrophysical models.  Unsupervised classification of GW candidates with convolutional neural networks have also been applied \citep{George18}. 

Gravitational waves are predicted to span a wide range of frequencies requiring different technologies to detect. Low-frequency gravitational waves are emitted from pairs of supermassive black holes in distant galaxies in tight orbits that power quasars and other active galactic nuclei, each of which is millions times more massive than those detected by LIGO.  
It is hoped these can be detected with a clever method using millisecond pulsars.  Radio telescopes have been observing Galactic millisecond pulsars, produced by spun-up rapidly rotating neutron stars, for decades but only recently have been harnessed for low-frequency GW detection \citep{Taylor16, Mingarelli17}.  Time series of pulsar pulse arrival times from pulsars distributed through the Milky Way Galaxy provide a Galaxy-scale detector for nanohertz GWs.  The astrophysical signals here are not short chirps but rather continuous periodic GWs emitted by pairs of orbiting supermassive black holes in distant galaxies. A stochastic autocorrelated GW noise from many orbiting black holes is expected.

Careful tracking and modeling of neutron star pulse arrival time give a very precise measurement of distance from Earth to the pulsar.  When a set of pulsars around the Galaxy are monitored in a pulsar timing array, it provides the ability to detect minute variations in space-time due to the passage of long wavelength GWs. Correlating the residuals from model predictions across pairs of pulsars leverages the common influence of a gravitational-wave background against unwanted, uncorrelated noise. So far, no astrophysical signal has been reported  \citep{Arzoumanian18}.  

The NAGOGrav time series data present considerable scientific, statistical and computational challenges.  This analysis starts with multicomponent, nonlinear model of a non-Gaussian and non-stationary time series \citep{Arzoumanian15, Arzoumanian18}.  A model of pulse arrival times must take into account: the astronomical characteristics of the pulsar (spin period, proper motion); gravitational effects from any stellar companions; time-varying dispersion measure variations from the propagation of the pulse through the inhomogeneous Galactic interstellar medium; autoregressive `red' noise of unknown origin with timescales of weeks to years; white noise (radio receiver noise and some additional component of unknown origin).  And finally, the model seeks a non-zero amplitude for a GW signal which gives a unique correlation pattern across multiple pulsars.  The GW component of the model typically rests on assumptions of an isotropic distribution of black holes and circular orbits of black hole binaries. Best-fit models are typically obtained using Bayesian inference with uninformative priors.

The models here, and in more broadly in astrostatistics, traditionally assume the non-GW behaviors are stationary and Gaussian.  But newer precise data are beginning to show traces of non-Gaussian and non-stationary features. Common analysis has not yet benefited from statistical methods such as multi-level hierarchical modeling, Gibbs sampling, or nonlinear time series modeling.  Additionally, testing for GW-induced distortions in this complicated model is often addressed with computationally expensive techniques involving Gaussian Processes regression and Bayesian inference with MCMC sampling \citep{Haasteren2014}. More efficient approaches are needed.

\section{TWO EXAMPLES OF MACHINE LEARNING IN ASTRONOMY} \label{ml.sec}

Machine learning techniques are growing exponentially in the astronomical literature covering a wide range of questions in planetary, stellar, extragalactic, time domain and cosmological fields. A new generation of young astronomers are quickly becoming proficient in their use. 

\subsection{Photometry of Blended Galaxies} \label{blend.sec}

The upcoming generation of deep and wide galaxy surveys from ground (LSST) and space (Euclid, WFIRST) observatories will present image processing challenges for which new methodologies are imperative \citep{Morgan2017}.  Of particular interest is the capability to extract photometric information overlapping, or blended, objects. As the instruments become more sensitive, the blending probability increases; half of the sources observed by LSST are predicted to have some level of overlap.  This blending can involve two galaxies that are physically interacting, two unrelated galaxies that overlap by chance, or chance superposition of a Galactic star with a distant galaxy. While astronomically motivated methods image segmentation algorithms are widely used \citep{SExtractor, Mancone2013}, interest has increased in data-driven approaches to address this problem \citep{Melchior2018}.

\begin{figure}[b]
    \centering
    \includegraphics[width=0.6\textwidth]{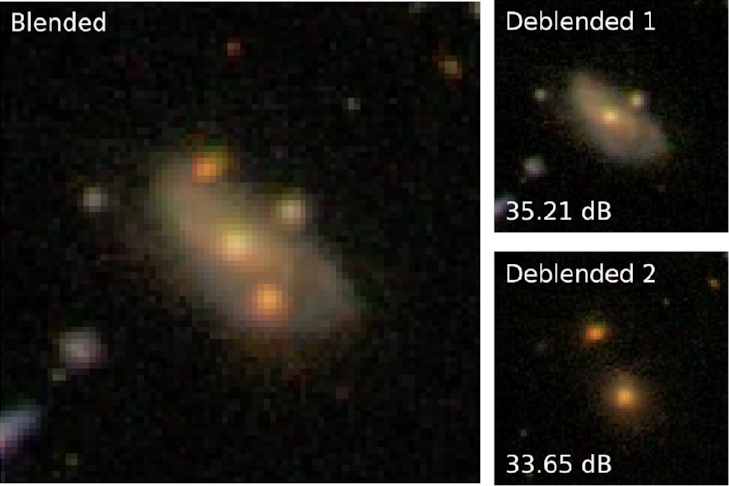}
    \caption{An artificial image constructed of two blended galaxy groups from the SDSS survey (left) and a successful decomposition using a branched generative adversarial network \citep{Reiman2019}.}
    \label{fig:blended}
\end{figure}

\begin{figure*}
\centering
\includegraphics[width=0.85\textwidth]{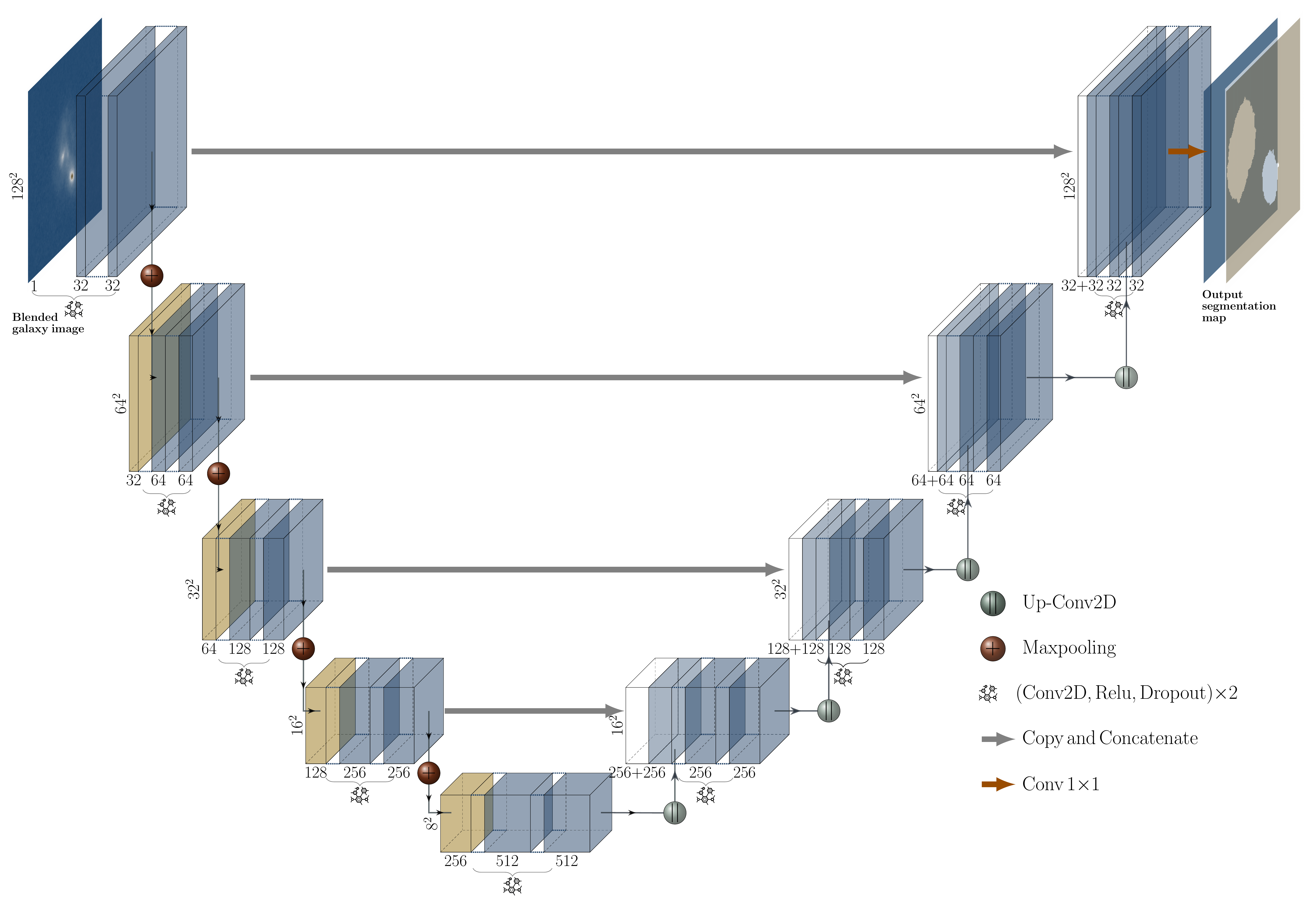}
\caption{A U-net network that takes as input an image of blended system and outputs two fractional segmentation maps, each pixel corresponding to the fraction of flux belonging to a given galaxy \citep{Boucaud2020}.}
\label{fig:unet}
\end{figure*}

Most recently, deep learning has been used to process galaxy images, particularly for the classification of galaxy morphologies \citep{Dieleman2015, Barchi2017, DominguezSanchez2018, HuertasCompany2018, Khalifa2018}. 
Techniques seek the recovery of galaxy features in noisy images with generative adversarial networks \citep{Schawinski2017}, the search for strong lensing effects with deep learning networks \citep{Lanusse2018}, and for deblending galaxies \citep[Figure~\ref{fig:blended},][]{Reiman2019}. \citet{Boucaud2020} implement a modular version of U-net architecture \citep{Ronneberger:2015aa} to recover the  fractional segmentation map; this is an image with pixel values between 0 and 1 estimating the fraction of flux belonging to a given galaxy.  Figure~\ref{fig:unet} displays the input blended galaxies on the top left and the recovered masks on the top right.  The method outperformed more traditional approaches, stressing the power of deep learning architecture tailored for astronomical image processing.  

\subsection{The Accelerating Expansion of the Universe} \label{SNIa.sec}

The expansion of the Universe since the Big Bang 13.7 billion years ago is naturally decelerated by the gravitational attraction of Dark Matter and ordinary matter. So it was a major surprise two decades ago when distance measurements of Type Ia supernovae (SNe Ia) provided the first empirical evidence that our Universe is experiencing an accelerated expansion.  SNe Ia are particularly useful because they behave as standard candles; that is, their intrinsic luminosities can be inferred enabling accurate estimates of their distances for a given cosmological model. Whilst the 
universal acceleration first detected by SNe Ia is corroborated by independent observations of the cosmic background radiation and other lines of evidence, SNe Ia continue to play a major role in our quest to understand Dark Energy, the unknown phenomena that causes this accelerated expansion.  Dedicated surveys are underway to improve the statistics of Dark Energy measurements; LSST is expected to measure $\sim 300,000$ SNe Ia light curves over the period of 10 years.  However, other types of SNe and transient objects confuse the situation.  Only a few percent of LSST SNe are expected to have spectroscopic information which provide accurate labels.  So the community must rely on sparse photometric light curves to classify transients and thus obtain SNe Ia samples. 

Ideally, the SNe photometric classification problem could be formulated as a case of straight-forward supervised learning. A subset of SNe with both spectroscopic and photometric information can give training sets for different SNe types.  This formulation of the problem was recently submitted to the scrutiny of the methodology community through the Kaggle PLAsTiCC competition \citep{plasticc}, which attracted more than 1000 participating teams. Nevertheless, a realistic treatment is more complex for several reasons. 

A particularly difficult characteristic of the SN classification problem is the discrepancy between spectroscopic and photometric samples, violating a central assumption underlying most learning algorithms.  Due to the heavy observational burden,  spectroscopy will not be available for the faintest LSST SNe, 
leading to biased training sets \citep{Childress17}. Several efforts are underway to treat this issue in machine learning-based analyses \citep{Revsbech2018}, including the use of deep learning. 

This leads to the problem of optimizing the distribution of spectroscopic resources to construct a training sample that maximizes accurate classification  with a minimum number of labels.  Within the framework of active learning, \citet{Ishida2019} provide observers with a spectroscopic follow-up protocol on a night-by-night basis.  The method iteratively identifies which objects in the target (photometric) sample would most likely improve the classifier if included in the training data - allowing sequential updates of the learning model with a minimum number of labeled instances. While the approach should outperform standard supervised learning models, caveats still remains. Better modeling of uncertainty quantification of the  photometric measurement errors, awareness of anomaly objects,  and a seamless inclusion of astronomical information into the loss function are needed.

\section{TWO CLASSIC PROBLEMS IN ASTROSTATISTICS}
\label{classic.sec}

While major projects are underway to advance machine learning methods for Big Data problems in astronomy, many research efforts can benefit from innovations in more classical statistical inference methodology.  

\subsection{Heteroscedastic Measurement Errors} \label{measerr.sec}

Heteroscedasticity occurs when errors are not uniform across a sample but exhibit dependencies on the measurement process or some (often unknown) properties of the objects under study.  In regression, the heteroscedasticity may be included within hierarchical statistical models as nuisance hyperparameters \citep{Carroll06}.   Heteroscedastic measurement errors are ubiquitous in astronomy but with a crucial difference:  astronomers measure the errors directly, so the variance of the error enters into the input data along with the property under study.   This is possible because astronomers build carefully calibrated instruments with known noise characteristics, and additional noise from the celestial environment or observational conditions can be measured simultaneously with the property of interest.  In imagery, it is the noise of the dark sky next to a star or galaxy; in spectroscopy, the noise of the spectrum next to an emission line of interest; and in time series, the temporal behavior before and after an event of interest.  It is thus standard procedure throughout astronomy for each property measured in each object to be accompanied by its own heteroscedastic measurement error with known variance.

The only common use of these measurement errors is in regression with weighted least squares which minimizes the sum $\Sigma_{i=1}^n  (x_i - M(\theta)_i)^2 / \sigma_i^2$.  Here $x_i$ is the observed property of interest for the $i$-th measurement, $M_i$ is the model prediction with parameters $\theta$, and $\sigma_i^2$ is the observed variance of the measurement.  Astronomers call this `minimum-$\chi^2$ regression’ hoping that the sum is chi-squared distributed.  The problem is that other components to the total error are often present which may be added in quadrature to the denominator of the sum.  The model is often misspecified in realistic situations so that the total error does not account for the sample variance and the sum is no longer chi-squared distributed.  

There is great need for statistical methodology treating known heteroscedastic measurement errors in astronomy beyond this regression problem.  First, in many cases the value of $x_i$ is not much greater than $\sigma_i$, and the object is considered to be undetected and the measurement $x_i$ is replaced by a left-censored value like $<  3 \times \sigma_i$.  The problem is that standard survival analysis treating censoring does not also treat heteroscedastic weighting of the clearly detected data points.   Second,  astronomers have other statistical needs such as cluster and classification which need to incorporate measurement errors.  Although this may be possible for parametric treatment like Gaussian mixture models, heteroscedastic errors are not easily adapted to  nonparametric algorithmic clustering and classification procedures.  

One approach to the problem is a partially Bayesian solution called the  iterative conditional modes which uses componentwise maximization routine to find the mode of the posterior \citep{Berry2002}. Another promising direction in the machine learning community is the development of a framework for Bayesian deep learning \citep{Yarin2016}. Similarly,  the Probably Approximately Correct Bayesian (PAC-Bayesian) Learning formalism has been used to  give theoretical guarantees to cutting edge topics in machine learning as deep learning and domain adaptation \citep{Guedj2019}. 

\subsection{Domain-Specific Probability Distributions}
\label{statdistr.sec}

Astronomers often seek analytic formulae to model the distributions of properties of scientific interest.  One of the most famous is the `Schechter function’ that accurately describes the distribution of galaxy luminosities with a truncated power law (Pareto) at low luminosities and an exponential distribution at high luminosities.  But astronomers did not realize this is just Euler’s gamma distribution.  A similar issue is the stellar IMF which resembles a power law at high masses (known as the 1955 `Salpeter function’) and a lognormal at low masses.  This is usually modeled as a piece-wise composite function such as Kroupa’s or Chabrier’s IMF. But \citet{Maschberger13} proposes a continuous generalized log-logistic distribution with three parameters, 
\begin{equation}
P_{L3}  (m) \propto  \frac{m}{\mu} ^{- \alpha} \left(1 + \left( \frac{m}{\mu} \right)^{1 - \alpha} \right)^{-\beta}.
\end{equation}

In a different area of astrophysical study, the distribution of particle energies in plasmas emerging from the Sun and measured throughout the Solar System is found to follow a `kappa distribution’ that is Maxwellian (Gaussian) at low energies and power law at high energies
\begin{equation}
f^\kappa (v) =  \frac{1}{(\pi \kappa v_\kappa^2)^{3/2}} \frac{\Gamma(\kappa+1)}{\Gamma(\kappa - 1/2} \left( 1 + \frac{v^2}{\kappa v_\kappa^2} \right) ^{-(\kappa + 1)}.
\end{equation}

Astrophysical insights can emerge from linking such distribution functions to generative stochastic processes.  For example, statisticians \citet{Reed04} show how a broad class of double-Pareto functions can be produced by killed multiplicative Brownian motion processes.  Astrophysicist \citet{Collier93} similarly shows that kappa distributions can arise from random walks in velocity governed by Lévy flight probability distributions.

\section{ASTROSTATISTICAL CHALLENGES FOR STATISTICIANS}
\label{statisticians.sec}

Contemporary astronomical data analysis often elude the capabilities of classical statistical techniques, and inevitably requires the use and development of sophisticated, and sometimes novel, statistical tools.

Astronomy requires expertise in vast fields of statistics and information science: nonparametric and parametric inference (especially Bayesian), high-dimensional nonlinear regression, censoring and truncation, measurement error theory, spatial point processes, image analysis, time series analysis, multivariate analysis, clustering and classification, and many other forms of machine learning. Statistical models range from simple heuristic power law regressions to high-dimensional non-linear models from astrophysical theory. Samples sizes range from a dozen to billions of objects. A hierarchy of problems with several layers of uncertainty is often involved; for example, a survey subject to flux limits is then filtered by morphology, subject to classification, and multivariate relationships are sought involving properties that may or may not be subject to the original truncation. When errors are non-Gaussian, analysis can benefit from generalized linear modeling  \citep{deSouza2015A,deSouza2015}.  

Clearly, cross-disciplinary collaboration and research is not only desired but imperative.  Yet most of the astrostatistical innovations reviewed here have been developed by astronomers who have little formal training in the mathematical and computational sciences. Complex astrostatistical procedures are often developed in isolation from the mainstream of methodological studies.  Education of astronomers in methodology is generally informal with  emphasis on Python-based software and hack weeks rather than thorough university-based courses.  Research in astronomy can be well-funded, but resources for methodological development are scarce; astrostatistical efforts are typically informally embedded within formal science and software projects. An institute or observatory will employ dozens of researchers with university degrees in astronomy and physics but very few with degrees in statistics, applied mathematics or computer science.  

Despite these structural constraints, a vibrant field of astrostatistics has emerged since the 1990s with international conferences, training workshops, and several cross-disciplinary scholarly organizations.   An informal online Facebook group on astrostatistics started in 2013 has grown to nearly 5000 members. The Cosmostatistics Initiative founded in 2014 created a successful interdisciplinary science development environment where innovative astrostatistics projects are developed and disseminated.  The Statistical and Applied Mathematical Sciences Institute has run several astrostatistical programs.  

In some respects, statisticians can readily enjoy the fruits of astronomical observations for statistical study: a vast range of data are freely available.  The astronomical community has a long-standing tradition, often legally binding, of making both raw and analyzed data accessible on the Web.  The U.S. National Aeronautical and Space Administration and European Space Agency operate large science archive centers for satellite observatories.  Ground-based data are available from the European Southern Observatory, U.S. National Optical Astronomy Observatory and National Radio Astronomy Observatory, and other major institutions.  The International Virtual Observatory Alliance provides an integrated interface to these and hundreds of other distributed databases.   The Vizier and SIMBAD services from France’s Centre de Données Astronomique, and the NASA/IPAC Extragalactic Database, provide convenient Web-based access to published tabular data.  The Smithsonian/NASA Astrophysics Data System provides superb bibliographic services with full-text, references and citations for nearly the entire astronomical research literature. 

But in other respects, conducting astrostatistical research requires considerable care.  The datasets are often subject to selection biases inaccessible to non-experts.  The scientific questions are often complex, so analyses must be designed within the context of established research programs.  Statistical studies in cosmology can be particularly challenging.  Analyses of cosmic microwave background maps need to be interpreted within the $\Lambda$CDM cosmological model, and three-dimensional cosmography of Dark Matter combines complexities of both statistical weak lensing signatures and the imprecision of photo-z distances.  

Interested statisticians can meet astronomers through newly formed scholarly societies and at specialized conferences.  The methodology community has the  International Statistical Institute’s Astrostatistics Special Interest Group, American Statistical Association’s Astrostatistics Interest Group, IEEE Task Force on AstroData Mining, and the independent International Astrostatistics Association and International AstroInformatics Association.  The astronomical community has the International Astronomical Union’s Commission on Astroinformatics and Astrostatistics, and the American Astronomical Society’s Working Group on Astroinformatics and Astrostatistics.   Ongoing conference series include {\it Statistical Challenges in Modern Astronomy}, {\it Astroinformatics}, and {\it Astronomical Data Analysis} as well as sessions at Joint Statistical Meetings, World Statistics Congresses, and American Astronomical Society meetings.   The large LSST project has an Information and Statistics Science Collaboration.   

The purpose of astronomy is to discover, characterize, and gain physical insight into cosmic phenomena.  Tremendous successes have emerged in the past century, propelled by amazingly sophisticated and sensitive instrumentation.  Astrostatistics plays an important integrative role, bridging raw data to intelligible information, and information to insightful astrophysical models.  Astrostatistics is a young, fertile field in which interdisciplinary communities can grow, providing essential expertise to further our understanding of the Universe we inhabit. 

{\it Acknowledgements:}  Astrostatistics at Penn State is supported by NSF grant AST-1614690, NASA grant 80NSSC17K0122, and the Eberly College of Science through the Center for Astrostatistics. EEOI is supported by a 2018-20 CNRS MOMENTUM fellowship. 

\bibliographystyle{ar-style1}
\bibliography{ref_Astrostatistics}

\end{document}